
 \def\eth{\hbox{$\partial$\kern-0.25em\raise0.6ex\hbox{\rm\char'40}}} \def\X{X}

\font\mbf=cmbx10 scaled\magstep1

\def\bs{\bigskip}
\def\ms{\medskip}

\def\ni{\noindent}
\def\cl{\centerline}

\def\ref#1#2#3#4{#1\ {\it#2\ }{\bf#3\ }#4\par}
\def\refb#1#2#3{#1\ {\it#2\ }#3\par}
\def\CQG{Class.\ Qu.\ Grav.}
\def\PR{Phys.\ Rev.}

\def\thorn{I\kern-0.4em\raise0.35ex\hbox{\it o}}
\def\I{\int_S{*}}
\def\O#1{\left.#1\right\vert_S}
\def\H#1{\left.#1\right\vert_H}
\def\half{{\textstyle{1\over2}}}

\magnification=\magstep1

\cl{\mbf Spin-coefficient form of the new laws of black-hole dynamics}
\bs\cl{\bf Sean A. Hayward}
\ms\cl{Department of Physics}
\cl{Kyoto University}
\cl{Kyoto 606-01}
\cl{Japan}
\ms\cl{Faculty of Mathematical Studies}
\cl{University of Southampton}
\cl{Southampton SO9 5NH}
\cl{United Kingdom}
\bs\cl{14th June 1994}
\bs\ni
{\bf Abstract.}
General laws of black-hole dynamics,
some of which are analogous to the laws of thermodynamics,
have recently been found for a general definition of black hole
in terms of a future outer trapping horizon,
a hypersurface foliated by marginal surfaces of a certain type.
This theory is translated here into spin-coefficient language.
Second law: the area form of a future outer trapping horizon
is generically increasing, otherwise constant.
First law: the rate of change of the area form
is given by an energy flux and the trapping gravity.
Zeroth law: the total trapping gravity of a compact outer marginal surface
has an upper bound, attained if and only if the trapping gravity is constant.
Topology law: a compact future outer marginal surface has spherical topology.
Signature law:
an outer trapping horizon is generically spatial, otherwise null.
Trapping law:
spatial surfaces sufficiently close to a compact future outer marginal surface
are trapped if they lie inside the trapping horizon.
Confinement law:
if the interior and exterior of a future outer trapping horizon
are disjoint, an observer inside the horizon cannot get outside.
\bs\cl{PACS: 04.70.Bw, 04.20.Cv, 04.20.Gz}
\bs\cl{Short title: Black-hole dynamics}
\bs\ni
{\bf 1. Introduction}
\ms\ni
The famous laws of black-hole dynamics
are formulated for stationary space-times [1],
or in the case of the second law, for asymptotically flat space-times [2].
These concern Killing horizons and event horizons respectively,
both of which define black holes in a special class of space-times.
Recently, general laws of black-hole dynamics have been found
for a general definition of black hole in arbitrary space-times,
the trapping horizon [3].
The results were obtained using a `2+2' formalism [4],
but can also be expressed in terms of spin-coefficients.
In this note, the new definitions and laws are given
in terms of the compacted spin-coefficient formalism [5].
The reader is referred to [3] for related results and a fuller discussion,
and would benefit from experience with Penrose \& Rindler [5],
particularly \S4.12--14.
Equation numbers in triples indicate the latter reference.
\ms
The basic objects are spatial 2-surfaces $S$ embedded in space-time,
and 3-surfaces $H$ foliated by such surfaces.
The signature of $H$ is not fixed,
so that one needs an approach allowing $H$ to be spatial, null or Lorentzian.
It is convenient to consider a double-null foliation,
namely two 1-parameter foliations of null 3-surfaces
intersecting in a 2-parameter foliation of spatial 2-surfaces.
Each $H$ determines a neighbouring double-null foliation
consisting of null 3-surfaces generated from each $S$
by light rays normal to $S$.
The double-null foliation is uniquely determined unless $H$ is itself null.
\ms
The geometry of double-null foliations is described in \S2
and particular aspects of the geometry of spatial 2-surfaces in \S5.
The main definitions are given in \S3
and various local, quasi-local and global properties of the definitions
are derived in \S4, \S6 and \S7 respectively.
\bs\ni
{\bf 2. Geometry of double-null foliations}
\ms\ni
The spin-coefficient formalism uses a basis of complex spin-vectors
$(o,\iota)$,
which determines a normalisation (2.5.46)
$$\chi=\{o,\iota\}\eqno(1)$$
and derivative operators (4.5.23)
$$(D,\delta,\delta',D')=
(\nabla_{o\bar o},
\nabla_{o\bar\iota},
\nabla_{\iota\bar o},
\nabla_{\iota\bar\iota})\eqno(2)$$
where $\nabla$ is the covariant derivative
and vectors are expressed as spin-vector dyads.
The spin-coefficients
$(\kappa,\rho,\sigma,\tau,\varepsilon,\alpha,\beta,\gamma,
\kappa',\rho',\sigma',\tau',\varepsilon',\alpha',\beta',\gamma')$
are defined by (4.5.26--27) as
$$\eqalignno{
D(o,\iota)&=(\varepsilon o-\kappa\iota,\gamma'\iota-\tau'o)\cr
\delta'(o,\iota)&=(\alpha o-\rho\iota,\beta'\iota-\sigma'o)\cr
\delta(o,\iota)&=(\beta o-\sigma\iota,\alpha'\iota-\rho'o)\cr
D'(o,\iota)&=(\gamma o-\tau\iota,\varepsilon'\iota-\kappa'o).&(3)\cr}
$$
The spin-basis also determines a basis of 1-forms (4.5.19)
$$(n',m,\bar m,n)=g(o\bar o,o\bar\iota,\iota\bar o,\iota\bar\iota)\eqno(4)$$
where $g$ is the space-time metric.
The identities
$$\eqalignno{
\chi\bar\chi\hbox{d}n
&=\kappa'm\wedge n'
+\bar\kappa'\bar m\wedge n'
+(\rho'-\bar\rho')\bar m\wedge m
+(\varepsilon'+\bar\varepsilon')n'\wedge n\cr
&\qquad+(\tau'-\bar\alpha'-\beta')m\wedge n
+(\bar\tau'-\alpha'-\bar\beta')\bar m\wedge n\cr
\chi\bar\chi\hbox{d}n'
&=\kappa\bar m\wedge n
+\bar\kappa m\wedge n
+(\rho-\bar\rho)m\wedge\bar m
+(\varepsilon+\bar\varepsilon)n\wedge n'\cr
&\qquad+(\tau-\bar\alpha-\beta)\bar m\wedge n'
+(\bar\tau-\alpha-\bar\beta)m\wedge n'&(5)\cr}
$$
can be derived by the Cartan method, as described in \S4.13 of [5].
Note that the corresponding expressions (4.13.44) are valid only if $\chi=1$.
\ms
If the null 3-surfaces are labelled by $\xi$ and $\xi'$,
then a suitable spin-basis is given by $n=\hbox{d}\xi$ and $n'=\hbox{d}\xi'$,
so that $(m,\bar m)$ spans the 2-surfaces of intersection.
This means that $n$ and $n'$ are closed, $\hbox{d}n=\hbox{d}n'=0$, so that
$$\eqalignno{
&0=\kappa=\rho-\bar\rho=\varepsilon+\bar\varepsilon=\tau-\bar\alpha-\beta\cr
&0=\kappa'=\rho'-\bar\rho'=\varepsilon'+\bar\varepsilon'
=\tau'-\bar\alpha'-\beta'.&(6)\cr}
$$
This constitutes a choice of gauge for the spin-coefficient formalism,
adapted to double-null foliations.
This gauge will be assumed henceforth.
Note that setting $\chi$ equal to 1
would be inconsistent with the gauge, in general.
Be warned that some of the formulas of [5] assume $\chi=1$,
and must be generalised accordingly.
\ms
Having fixed a particular double-null foliation,
there is still considerable gauge freedom entangled with the true geometry
in the spin-coefficients.
This is disentangled in the compacted spin-coefficient formalism,
\S4.12 of [5],
which separates out quantities with simple scaling behaviour
under rescalings of the spin-basis,
described as weighted scalars.
Additionally, the derivative operators $(D,D',\delta,\delta')$
are replaced by weighted operators $(\thorn,\thorn',\eth,\eth')$
defined by (4.12.15).
This allows the Einstein equations to be written without the occurrence of
the unweighted spin-coefficients $(\varepsilon,\alpha,\beta,\gamma)$ and
$(\varepsilon',\alpha',\beta',\gamma')$, (4.12.32).
The remaining spin-coefficients
$(\rho,\sigma,\tau)$ and $(\rho',\sigma',\tau')$,
the normalisation $\chi$,
and the operators $(\thorn,\thorn',\eth,\eth')$
have weights given by (4.12.13), (4.12.14) and (4.12.17).
These quantities will be the ones used.
In particular, the following quantities are zero-weighted,
which means that they are scalar invariants of the double-null foliation:
$\rho\rho'/\chi\bar\chi$,
$\sigma\sigma'/\chi\bar\chi$,
$\tau\bar\tau/\chi\bar\chi$,
$\eth'\tau/\chi\bar\chi$,
$\thorn'\rho/\chi\bar\chi$.
\ms
Before moving on, there is one potential source of confusion to be removed,
namely the difference between invariants of a double-null foliation
at a given $S$ and invariants of a 2-surface $S$ considered in isolation.
The latter class is smaller than the former,
since there are many different double-null foliations based on $S$.
In particular, the normalisation $\O\chi$ could be fixed by coordinate choice
if the double-null foliation were not fixed.
Since $\tau+\bar\tau'=(\chi\bar\chi)^{-1}\delta(\chi\bar\chi)$,
this would also fix $\O{(\tau+\bar\tau')}$.
Of the weighted spin-coefficients, this would leave
$\O{(\rho,\rho',\sigma,\sigma',\tau-\bar\tau')}$ as free data.
Specifically,
$\O{\tau\bar\tau/\chi\bar\chi}$,
$\O{\eth'\tau/\chi\bar\chi}$ and
$\O{\thorn'\rho/\chi\bar\chi}$
are invariants of a double-null foliation at $S$,
but not of a 2-surface $S$ in isolation.
\bs\ni
{\bf 3. Marginal surfaces, trapping horizons and trapping gravity}
\ms\ni
The space-time is assumed to be time-orientable,
and the null vectors dual to $n$ and $n'$ are taken to be future-pointing.
Equivalently, $\thorn$ and $\thorn'$ differentiate to the future.
\ms\ni
{\it Definitions} [3].
A {\it marginal surface} is a spatial 2-surface $S$
on which one convergence vanishes, fixed henceforth as $\O\rho=0$.
A {\it trapping horizon} is the closure $\overline H$ of a 3-surface $H$
foliated by marginal surfaces on which
$\H{\rho'}\not=0$ and $\H{\thorn'\rho}\not=0$,
where the double-null foliation is adapted to the marginal surfaces.
The trapping horizon and marginal surfaces are said to be
{\it outer} if $\H{\thorn'\rho}>0$,
{\it inner} if $\H{\thorn'\rho}<0$,
{\it future} if $\H{\rho'}>0$ and {\it past} if $\H{\rho'}<0$.
\ms\ni
In the example of the Kerr or Reissner-Nordstr\"om black hole,
the future event horizon is a future outer trapping horizon,
the future Cauchy horizon is a future inner trapping horizon,
the past event horizon is a past outer trapping horizon,
and the past Cauchy horizon is a past inner trapping horizon.
The idea behind a future outer trapping horizon is simply that
the ingoing light rays should be converging, $\H{\rho'}>0$,
and the outgoing light rays should be instantaneously parallel on the horizon,
$\H\rho=0$,
and diverging just outside the horizon and converging just inside,
$\H{\thorn'\rho}\ge0$.
The degenerate cases, where $\thorn'\rho$ or $\rho'$ vanishes somewhere on $H$,
have been excluded since they include pathological cases
which do not indicate the existence of a black hole.
The existence of a future outer trapping horizon
provides a general definition of a black hole.
Slightly more generally,
one may define a {\it trapping boundary} [3]
as a boundary of an inextendible trapped region,
namely a region in which each point lies on a trapped surface.
Assuming technical smoothness and uniformity conditions,
a trapping boundary must be
either a future or past, outer or inner trapping horizon,
or a degenerate horizon [3].
\ms\ni
{\it Definition} [3].
The {\it trapping gravity} of an outer trapping horizon $\overline H$ is
$$\kappa=\H{\sqrt{\half\thorn'\rho/\chi\bar\chi}}\eqno(7)$$
using the previously liberated symbol $\kappa$.
According to the previous section, $\kappa$ is an invariant of $H$
unless $H$ is null.
Trapping gravity is an analogue of surface gravity for a trapping horizon,
measuring the strength of the trapping effect at the horizon.
Note that $\kappa>0$, which may be relaxed to $\kappa\ge0$
if one wishes to include degenerate trapping horizons.
The putative {\it third law}
expresses the physical inattainability of degenerate trapping horizons.
The possibility of a trapping horizon developing a degeneracy
is implicitly allowed in the definition,
since extending $\kappa$ to $\overline{H}$ leads only to
$\kappa|_{\overline{H}}\ge0$.
\bs\ni
{\bf 4. Local laws}
\ms\ni
The local laws,
namely those which refer to a single point of the trapping horizon,
use the field equation (4.12.32a),
$$\thorn\rho=\rho^2+\sigma\bar\sigma+\Phi_{00}\eqno(8)$$
referred to as the {\it focussing equation},
and either the weak energy condition
or the null energy (or convergence) condition,
which imply
$$\Phi_{00}\ge0.\eqno(9)$$
\ms\ni
{\it Signature law} [3].
If the null energy condition holds on a trapping horizon:
the horizon is null if and only if $\sigma$ and $\Phi_{00}$ vanish;
otherwise, an outer trapping horizon is spatial
and an inner trapping horizon is Lorentzian.
\ms\ni
{\it Proof.}
Let $\Delta=\H{b\thorn}-\H{c\thorn'}$ be the derivative,
unique up to magnitude,
which generates the marginal surfaces foliating $H$.
Then $\H{\Delta\rho}=0$,
so that $c/b=\H{\thorn\rho/\thorn'\rho}$.
The focussing equation yields $\thorn\rho\ge0$,
so that $c/b\ge0$ for an outer trapping horizon
and $c/b\le0$ for an inner trapping horizon,
with $c/b|_p=0$ at $p\in H$
if and only if $\sigma|_p=0$ and $\Phi_{00}|_p=0$.
The result follows from the fact that the direction given by $\Delta$
is spatial, null or temporal
as $c/b$ is positive, zero or negative respectively.
\ms\ni
{\it Second law} [3].
If the null energy condition holds on a trapping horizon,
the area form of a future outer or past inner trapping horizon
is non-decreasing,
and the area form of a past outer or future inner trapping horizon
is non-increasing,
in all cases being constant only for null trapping horizons.
\ms\ni
{\it Proof.}
Note that $\thorn{*}1=-{*}2\rho$
and $\thorn'{*}1=-{*}2\rho'$,
where $*$ denotes the Hodge operator of $S$,
so that $*1$ is the area 2-form.
Then $\H{\Delta{*}1}=\H{*2c\rho'}$.
Fixing the orientation of $\Delta$ by $b>0$ gives
$\H{c\rho'}\ge0$ for a future outer or past inner trapping horizon,
and $\H{c\rho'}\le0$ for a past outer or future inner trapping horizon,
with $c\rho'|_p=0$ if and only if the trapping horizon is null at $p\in H$.
\ms\ni
{\it Corollary.}
The area form of a future trapping horizon
is increasing, constant or decreasing
as the horizon is spatial, null or Lorentzian respectively.
\ms\ni
{\it First law} [3].
If the null energy condition holds
on an outer trapping horizon $\overline{H}$,
$$\kappa\Delta{*}1
=\H{*{\rho'\sqrt{\Phi_{00}+\sigma\bar\sigma}\over{\chi\bar\chi}}}\eqno(10)$$
where $\Delta$ has unit normalisation if spatial.
\ms\ni
{\it Proof.}
If the trapping horizon is null, $\Delta{*}1=0$.
If it is spatial, $\Delta$ may be normalised by
$\Delta=\H{((2c\chi\bar\chi)^{-1}\thorn-c\thorn')}$,
so that
$2c^2=\H{\thorn\rho/\chi\bar\chi\thorn'\rho}
=\H{(\Phi_{00}+\sigma\bar\sigma)/2(\chi\bar\chi\kappa)^2}$
and $\Delta{*}1=\H{*2c\rho'}
=\H{*\rho'\sqrt{\Phi_{00}+\sigma\bar\sigma}/\chi\bar\chi\kappa}$.
\ms\ni
An equivalent first law and some similar ideas have appeared elsewhere [6],
but using a `temperature' $\thorn'\rho/2\pi\rho$
which is not invariant under rescaling of the null normals.
\bs\ni
{\bf 5. Geometry of spatial 2-surfaces}
\ms\ni
Introduce the complex curvatures
$$\eqalignno{
&K=\sigma\sigma'-\rho\rho'-\Psi_2+\Phi_{11}+\Pi\cr
&K^*=\tau\tau'-\Psi_2-\Phi_{11}+\Pi&(11)\cr}
$$
where $K$ is defined as in (4.14.20),
and $K^*$ is dual to $K$ under the Sachs operation (4.12.47).
These quantities occur in the commutators
$$\eqalignno{
&(\eth\eth'-\eth'\eth)\eta=-(K+\bar K)\eta\cr
&(\thorn\thorn'-\thorn'\thorn)\eta=(K^*+\bar K^*)\eta
+(\bar\tau-\tau')\eth\eta+(\tau-\bar\tau')\eth'\eta&(12)\cr}
$$
obtained from (4.12.33--35),
where $\eta$ is a scalar of weight $\{1,-1\}$.
The first of these commutators, (4.14.19),
shows that $(K+\bar K)/\chi\bar\chi$ is the Gaussian curvature of $S$,
slightly generalising (4.14.21).
For compact orientable $S$, there is the Gauss-Bonnet theorem,
slightly generalising (4.14.42),
$$\I{K+\bar K\over{\chi\bar\chi}}=4\pi(1-g)\eqno(13)$$
where $g$ is the genus of $S$.
Also,
$\I\eth'\tau/\chi\bar\chi=0$
by the Gauss divergence theorem (4.14.69),
since $\eth'\tau/\chi\bar\chi=\eth'(\tau/\chi\bar\chi)$ is a total divergence.
The imaginary part of the cross-focussing equation then yields
$\I(K-\bar K)/\chi\bar\chi=0$,
slightly generalising (4.14.43).
If $S$ is not necessarily compact or orientable, there are analogous results,
provided that
$(K+\bar K)/\chi\bar\chi$, $\eth'\tau/\chi\bar\chi$ and $\tau/\chi$
are integrable over $S$.
A marginal surface $S$ with this property
will be described as {\it well adjusted}.
Integrability of $\eth'\tau/\chi\bar\chi$ and $\tau/\chi$ ensures [7] that
$$\I{\eth'\tau\over{\chi\bar\chi}}=0\eqno(14)$$
whence the imaginary part of the cross-focussing equation yields
$$\I{K-\bar K\over{\chi\bar\chi}}=0.\eqno(15)$$
Integrability of the Gaussian curvature $(K+\bar K)/\chi\bar\chi$
gives the Cohn-Vossen inequality [7],
$$\I{K+\bar K\over{\chi\bar\chi}}\le2\pi\X\eqno(16)$$
where $\X$ denotes the Euler-Poincar\'e characteristic,
with $\X=2$ for a sphere, $\X=1$ for a plane or projective plane,
$\X=0$ for a torus, cylinder, Klein surface or M\"obius band,
and $\X\le-1$ for any other 2-manifold.
For a compact orientable 2-manifold,
the Cohn-Vossen inequality reduces to an equality,
recovering the Gauss-Bonnet theorem.
\bs\ni
{\bf 6. Quasi-local laws}
\ms\ni
The field equation (4.12.32f),
$$\thorn'\rho-\eth'\tau=\rho\rho'+\sigma\sigma'-\tau\bar\tau-\Psi_2-2\Pi
\eqno(17)$$
may be rewritten as
$$\thorn'\rho=\eth'\tau+2\rho\rho'-\tau\bar\tau+K-\Phi_{11}-3\Pi\eqno(18)$$
which is referred to as the {\it cross-focussing equation}.
The quasi-local laws, namely those which refer to the whole marginal surface,
use this equation and the dominant energy condition [2,5], which implies
$$\Phi_{11}+3\Pi\ge0.\eqno(19)$$
\ms\ni
{\it Topology law} [3].
If the dominant energy condition holds
on a well adjusted, future or past, outer marginal surface,
then it has spherical or planar topology.
\ms\ni
{\it Proof.}
The cross-focussing equation yields $0<\O{(\eth'\tau+K)/\chi\bar\chi}$
on an outer marginal surface $S$.
Integrating over well adjusted $S$ yields $\X>0$.
Future or past marginal surfaces must be orientable,
due to the differing signs of $\O\rho$ and $\O{\rho'}$,
so the topology is spherical, $\X=2$, or planar, $\X=1$.
\ms\ni
{\it Corollary}.
If the dominant energy condition holds
on a compact, future or past, outer marginal surface,
then it has spherical topology.
\ms\ni
{\it Definitions.}
For compact $S$, one may define the area $A$, irreducible energy $m$ [8],
angular energy $a$ [3] and material energy $q$ [3] by
$$\eqalignno{
&A=\I1\cr
&m=\sqrt{A/16\pi}\cr
&a^2={m^2\over{2\pi}}\I{\tau\bar\tau\over{\chi\bar\chi}}\cr
&q^2={m^2\over{2\pi}}\I{\Phi_{11}+3\Pi\over{\chi\bar\chi}}.&(20)\cr}
$$
The Hamiltonian energy [9] of $S$ is the real part of
$$E
={m\over{2\pi}}\I{K+\rho\rho'-\sigma\sigma'+\tau\tau'\over{\chi\bar\chi}}
={m\over{2\pi}}\I{K^*+2\Phi_{11}\over{\chi\bar\chi}}.\eqno(21)$$
If $\O\chi=1$,
the real part of the integrand is essentially the Hamiltonian density
which generates the double-null form of the vacuum Einstein equations [4].
Thus $(K^*+\bar K^*)/4\pi\chi\bar\chi$ may be interpreted as an energy density.
Whether the imaginary part $(K^*-\bar K^*)/4\pi\chi\bar\chi$ of the integrand
has any dynamical significance is unclear.
\ms\ni
{\it Zeroth law} [3].
If the dominant energy condition holds
on a compact, future or past, outer marginal surface $S$,
the total trapping gravity is bounded above by
$$\I\kappa\le4\pi\sqrt{m^2-a^2-q^2}\eqno(22)$$
with equality if and only if $\kappa$ is constant on $S$.
\ms\ni
{\it Proof.}
The cross-focussing equation, divided by $2\chi\bar\chi$, integrates to
$$\I\kappa^2=\pi\left(1-{a^2\over{m^2}}-{q^2\over{m^2}}\right)\eqno(23)$$
recalling that $S$ must have spherical topology,
and that $\I K/\chi\bar\chi=2\pi$.
The result follows from the Cauchy-Schwarz inequality,
$(\I f)^2\le A\I f^2$,
with equality if and only if $f$ is constant on $S$.
\ms\ni
Tod [10] has remarked that
$(4\pi q)^2=A\I|\varphi_{01}/\chi|^2$ in the Einstein-Maxwell case,
which is related to the charge $Q$ defined by (6.4.4),
$4\pi Q=\I|\varphi_{01}/\chi|$.
In particular, $Q^2\le q^2$,
with equality if and only if
the charge density $|\varphi_{01}/4\pi\chi|$ is constant on $S$.
Thus there is also the bound $\I\kappa\le4\pi\sqrt{m^2-a^2-Q^2}$.
\ms\ni
{\it Cosmic area limit} [11].
If there is a cosmological constant $\lambda>0$
and the dominant energy condition holds
on a compact, future or past, outer marginal surface $S$,
the area of $S$ is bounded above by
$$A\le4\pi/\lambda.\eqno(24)$$
\ms\ni
{\it Proof.}
Adding $\lambda$ to the cross-focussing equation yields
$$\thorn'\rho=\eth'\tau+2\rho\rho'-\tau\bar\tau+K-\Phi_{11}-3\Pi-\half\lambda
\eqno(25)$$
which integrates to
$$A\lambda=4\pi\left(1-{a^2\over{m^2}}-{q^2\over{m^2}}\right)
-4\I\kappa^2\eqno(26)$$
so that $A\lambda\le4\pi$.
\ms\ni
Finally, for compact $S$, the second law integrates to
$$\Delta A\ge0\eqno(27)$$
for a future outer trapping horizon,
and the first law may also be integrated to
$$\Delta A=
\I{\rho'\sqrt{\Phi_{00}+\sigma\bar\sigma}\over{\chi\bar\chi\kappa}}.
\eqno(28)$$
\bs\ni
{\bf 7. Global laws}
\ms\ni
A possible statement of the putative third law is that
if the trapping gravity $\kappa$ is non-zero on a marginal surface $S$,
$\kappa$ cannot become zero
on the trapping horizon $\overline{H}$ containing $S$,
or that such an occurrence is non-generic or unstable in some sense.
This is a conjecture about the global nature of trapping horizons,
which awaits further study.
Currently known properties of the space-time away from a trapping horizon
are given below.
\ms\ni
{\it Definitions.}
A compact orientable spatial 2-surface $S$ is {\it trapped} [12] if
$\O{\rho\rho'}>0$
and {\it mean convex} [13] if
$\O{\rho\rho'}<0$.
A trapped surface is {\it future} if $\O\rho>0$
and {\it past} if $\O\rho<0$ [3].
\ms\ni
{\it Trapping law} [3].
Given a compact marginal surface $S$
lying in a future or past, outer or inner trapping horizon,
any spatial 2-surface sufficiently close to $S$ in the double-null foliation
is trapped if it lies to one side of the horizon,
and mean convex if it lies to the other side.
The trapped surfaces are future (respectively past) trapped
for a future (respectively past) trapping horizon,
and lie to the future (respectively past) of an outer trapping horizon.
\ms\ni
{\it Proof.}
Consider any 3-surface $\Sigma$ which intersects $S$,
is not tangent to the trapping horizon at $S$,
and is foliated by spatial 2-surfaces
in the double-null foliation based on the marginal surfaces.
The foliation of $\Sigma$ is generated by a derivative $\Upsilon$
such that $\O{\Upsilon}=\O{(r\Delta+s\thorn')}$,
where $s>0$ fixes the orientation of $\Upsilon$.
Since $\O{\Upsilon\rho}=\O{s\thorn'\rho}$,
$\O{\Upsilon\rho}>0$ for an outer marginal surface
and $\O{\Upsilon\rho}<0$ for an inner marginal surface.
In either case, there are neighbourhoods $U_+$ and $U_-$ of $\Sigma$,
contiguous at $S$, such that $\rho|_{U_+}>0$ and $\rho|_{U_-}<0$.
For an outer trapping horizon, which is spatial or null,
$U_+$ lies to the future and $U_-$ to the past.
Also $\O{\rho'}>0$ (respectively $\O{\rho'}<0$)
for a future (respectively past) marginal surface,
so that one can choose $U_\pm$ such that
$\rho'|_{U_\pm}>0$ (respectively $\rho'|_{U_\pm}<0$).
This gives the result,
recalling that a future or past marginal surface is necessarily orientable.
\ms\ni
This gives a trapping horizon a preferred orientation,
with neighbouring trapped surfaces existing to one side of the horizon only.
Obtaining this property is the reason for the sign restrictions
$\H{\rho'}\not=0$ and $\H{\thorn'\rho}\not=0$
in the definition of trapping horizon.
If either sign varies over a marginal surface $S$,
varying $S$ will not generate neighbouring trapped surfaces.
\ms\ni
{\it Definitions} [14].
A 3-surface is {\it separating}
if its complement in space-time has two disjoint components.
For a separating trapping horizon,
the component containing the neighbouring trapped surfaces is
{\it inside} the horizon,
and the other component {\it outside} the horizon.
\ms\ni
{\it Confinement law} [14].
If the null energy condition holds on a separating outer trapping horizon,
an observer inside a future horizon cannot get outside,
and an observer outside a past horizon cannot get inside.
\ms\ni
{\it Proof.}
For a separating horizon,
any path between inside and outside must cross the horizon.
Since an outer trapping horizon is spatial or null,
an observer may cross the horizon in one direction only,
from outside to inside across a future trapping horizon
and from inside to outside across a past trapping horizon.
\bs\ni
Acknowledgements.
It is a pleasure to thank Paul Tod and James Vickers for discussions,
and Y.~Oshiro, K.~Nakamura and A.~Tomimatsu for pointing out reference [6].
This research was supported by the Japan Society for the Promotion of Science.
\bs
\begingroup
\parindent=0pt\everypar={\global\hangindent=20pt\hangafter=1}\par
{\bf References}\ms
\refb{[1] Carter B 1973 in}{Black Holes}
{ed: DeWitt C \& DeWitt B S (Gordon \& Breach)}
\refb{[2] Hawking S W 1973 in}{Black Holes}
{ed: DeWitt C \& DeWitt B S (Gordon \& Breach)}
\ref{[3] Hayward S A 1994}{General laws of black-hole dynamics, \PR}{D49}
{(in press)}
\ref{[4] Hayward S A 1993}\CQG{10}{779}
\refb{[5] Penrose R \& Rindler W 1986 \& 1988}
{Spinors and Space-Time Volumes 1 \& 2}{(Cambridge University Press)}
\ref{[6] Collins W 1992}\PR{D45}{495}
\ref{[7] Newman R P A C 1987}\CQG4{277}
\ref{[8] Christodoulou D \& Ruffini R 1971}\PR{D4}{3552}
\ref{[9] Hayward S A 1994}\PR{D49}{831}
\refb{[10] Tod K P 1994 (private communication)}{}{}
\ref{[11] Hayward S A, Shiromizu T \& Nakao K 1994}\PR{D49}{5080}
\refb{[12] Penrose R 1968 in}{Battelle Rencontres}
{ed: DeWitt C M \& Wheeler J A (Benjamin)}
\refb{[13] Hayward S A 1994}{Quasi-localisation of Bondi-Sachs energy loss}
{(gr-qc/9405071)}
\refb{[14] Hayward S A 1994}{Confinement by black holes}{(gr-qc/9405055)}
\endgroup
\bye